\documentclass[a4paper,fleqn,usenatbib]{mnras}

\usepackage[T1]{fontenc}
\usepackage{ae,aecompl}

\usepackage{graphicx}	
\usepackage{amsmath}	
\usepackage{amssymb}	

\title[NGC5252: Dual radio-emitting AGN?]{NGC~5252: a pair of radio-emitting active galactic nuclei?}

\author[X.-L. Yang et al.]{Xiaolong Yang,$^{1,2}$
Jun Yang,$^{3,4}$\thanks{E-mail: jun.yang@chalmers.se}
Zsolt Paragi,$^{5}$
Xiang Liu,$^{1,6}$
Tao An,$^{4,6}$
\and
Stefano Bianchi,$^{7}$
Luis C. Ho,$^{8,9}$
Lang Cui,$^{1,6}$
Wei Zhao$^{4,6}$
and Xiaocong Wu$^{4,6}$
\\
\\
$^{1}$Xinjiang Astronomical Observatory, Chinese Academy of Sciences, 150 Science 1-Street, 830011 Urumqi, P.R. China\\
$^{2}$University of Chinese Academy of Sciences, 100049 Beijing, P.R. China \\
$^{3}$Department of Earth and Space Sciences, Chalmers University of Technology, Onsala Space Observatory, SE-439\,92 Onsala, Sweden \\
$^{4}$Shanghai Astronomical Observatory, Chinese Academy of Sciences, 200030 Shanghai, P.R. China \\
$^{5}$Joint Institute for VLBI ERIC (JIVE), Postbus~2, NL-7990\,AA Dwingeloo, the Netherlands \\
$^{6}$Key Laboratory of Radio Astronomy, Chinese Academy of Sciences, 210008 Nanjing, P.R. China \\
$^{7}$Dipartimento di Matematica e Fisica, Universit\`a degli Studi Roma Tre, via della Vasca Navale 84, I-00146 Roma, Italy \\
$^{8}$Kavli Institute for Astronomy and Astrophysics, Peking University, Beijing 100871, P.R. China \\
$^{9}$Department of Astronomy, School of Physics, Peking University, Beijing 100871, P.R. China \\
}

\date{Accepted XXX. Received YYY; in original form ZZZ}

\pubyear{2016}

\begin{document}
\label{firstpage}
\pagerange{\pageref{firstpage}--\pageref{lastpage}}
\maketitle

\begin{abstract}
{The X-ray source CXO J133815.6$+$043255 has counterparts in the UV, optical, and radio bands. Based on the multi-band investigations, it has been recently proposed by \citet{kim15} as a rarely-seen off-nucleus ultraluminous X-ray (ULX) source with a black hole mass of $\geq10^4~\mathrm{M}_{\odot}$ in the nearby Seyfert galaxy NGC~5252. To explore its radio properties at very high angular resolution, we performed very long-baseline interferometry (VLBI) observations with the European VLBI Network (EVN) at 1.7~GHz. We find that the radio counterpart is remarkably compact among the known ULXs. It does not show a resolved structure with a resolution of a few milliarcsecond (mas), and the total recovered flux density is comparable to that measured in earlier sub-arcsecond-resolution images. The compact radio structure, the relatively flat spectrum, and the high radio luminosity are consistent with a weakly accreting supermassive black hole in a low-luminosity active galactic nucleus. The nucleus of NGC~5252 itself has similar radio properties. We argue that the system represents a relatively rare pair of active galactic nuclei, where both components emit in the radio.}

\end{abstract}




\section{Introduction}
\label{sec1}

Ultra-luminous X-ray sources (ULXs) are off-nucleus, point-like sources with an isotropic X-ray luminosity $L_\mathrm{X}\gtrsim3\times10^{39}$~erg\,s$^{-1}$. The high luminosity of ULXs implies super-Eddington accretion, a strong beaming effect or a non-stellar-mass black hole (BH). ULXs with $L_\mathrm{X}\gtrsim5\times10^{40}$~erg\,s$^{-1}$ are the most challenging to explain because the implied mass places them in the regime of the long-sought intermediate-mass BH class, $10^2\mathrm{M}_\odot<M_\mathrm{BH}<10^5\mathrm{M}_\odot$ \citep[e.g.,][]{Cropper+04, MFM04a, MFM04b, PDM04}. Intermediate-mass BHs are particularly important to our understanding of the formation of supermassive BHs \citep[see][and references therein]{MC04}. Currently, there are only a few promising intermediate-mass BH candidates, such as ESO 243$-$49 \citep[HLX--1:][]{cseh15a, far09, webb12}, M82 X--1 \citep{Dewangan06, Matsumoto01, pas14, SM03} and NGC 5408 X--1 \citep{DeMarco13, SM09}. Most of the ULXs likely have stellar-mass counterparts, as was shown in the cases of M\,101 ULX--1 \citep{liu13}, M\,31 ULX \citep{mid13}, NGC\,7793 \citep{Motch14} and M82\,X--2, the latter containing a neutron star \citep{bac14}.

Some ULXs in nearby galaxies have been detected in the radio. Most ULX radio counterparts have a bubble-like morphology on arcsecond scales \citep[e.g.,][]{cseh12}, likely inflated by jets of stellar-mass BHs. Recent, radio studies have revealed transient jets associated with X-ray outbursts of the central objects in ESO~243$-$49 \citep{webb12} and Holmberg~II X-1 \citep{cseh15b}. Compared with stellar-mass objects, intermediate-mass BHs are expected to have more luminous compact radio emission and thus may be detectable in nearby galaxies \citep{par06, kaa09}.

Low-luminosity AGNs (LLAGNs), powered by supermassive BH ($M_\mathrm{BH}>10^6\mathrm{M}_\odot$), are also potential objects hiding in the off-nucleus ULX sample. This is because LLAGNs have X-ray luminosities of $\sim10^{40}$ -- $10^{41}$~erg\,s$^{-1}$ (and lower; see Ho 2008, 2009), comparable to those of ULXs. Finding such an LLAGN associated with a ULX will provide direct evidence for a merger with the host galaxy, an important step in forming galaxies and supermassive BHs \citep[e.g.,][]{vol03}. A steady radio jet structure compact on mas scales is expected in hard-state LLAGNs at low accretion rates \citep[e.g.,][and references therein]{Paragi14}. So, very long-baseline interferometry (VLBI) observations of these sources will help to confirm LLAGN activity, in particular in cases where optical and X-ray evidence are missing \citep[e.g.,][]{par16}.

The source CXO~J133815.6$+$043255 has been recently reported by \citet{kim15} to have an X-ray luminosity of $L_\mathrm{X}=1.5\times10^{40}$~erg\,s$^{-1}$ at $z=0.022$ in the Seyfert galaxy NGC~5252 and an accreting black hole with a mass of $\geq10^4~\mathrm{M}_{\odot}$. The source is about $22\arcsec$ (about 10~kpc) away from the nucleus of NGC 5252. Compared to known ULXs, CXO~J133815.6$+$043255 is quite unusual. It has clear counterparts in the UV, optical, and radio bands and show some optical spectral properties similar to LLAGNs \citep{kim15}. In earlier Very Large Array (VLA) observations, the radio counterpart was clearly detected with a flux density of $3.2\pm0.2$~mJy at 1.4~GHz \citep{wil94} and $1.6\pm0.2$~mJy at 8.4~GHz \citep{kuk95}. With a resolution up to $0\farcs3$, it is still unresolved in the VLA observations at 8.4~GHz \citep{wil94} and in the Multi-Element Radio Linked Interferometer Network (MERLIN) observations at~1.6 GHz \citep{the01}. Alternatively, if it is a LLAGN, resembling an ULX due to its point-like morphology, all these characteristics would be quite natural. To find the most direct evidence, i.e. a persistent compact radio jet, for the LLAGN scenario, we performed European VLBI Network (EVN) observations.

In the Letter, we present a new dual radio-emitting AGN candidate. We introduce our EVN observations and data reduction in Section \ref{sec2} and show the VLBI imaging results of both NGC 5252 and CXO J133815.6$+$043255 in Section \ref{sec3}. In Section \ref{sec4}, we argue that the latter is a supermassive BH-powered LLAGN. We summarise our conclusions in Section \ref{sec5}.

\begin{figure}
\centering
\includegraphics[width=0.45\textwidth]{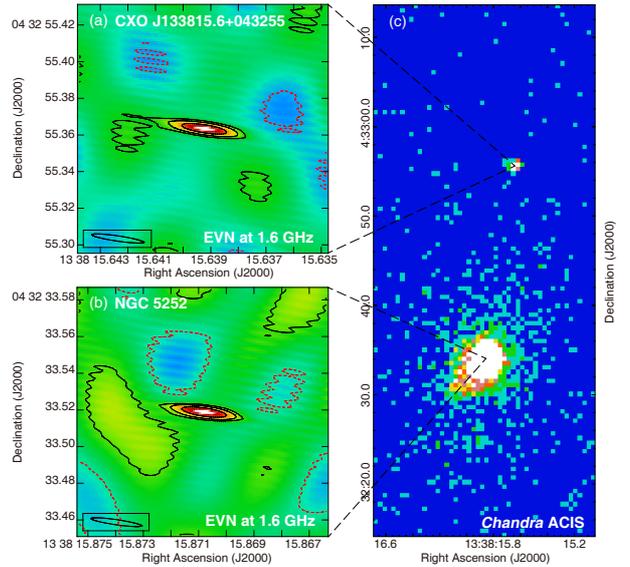}
\caption{Left: the EVN detections of CXO J133815.6$+$043255 and NGC~5252. Right: \textit{Chandra} ACIS-S image of the pair of sources in the 0.3~--~8 keV band (ObsId 15022). Natural weighting was used in the EVN images. The contours increase by a factor of $-1$, 1, 2, 4, and 8 from 3 times the off-source noise level, $0.05$~mJy\,beam$^{-1}$ for CXO J133815.6$+$043255 and $0.1$~mJy\,beam$^{-1}$ for NGC~5252. The beam has a full width to half maximum (FWMH) of $28\times2.9$~mas at position angle of 82$\degr$.} \label{fig1}
\end{figure}

\section{Observations and data reduction}
\label{sec2}

We observed both CXO J133815.6$+$043255 and NGC~5252 with the EVN at 1.66~GHz for three hours on 2015 December 2 (project code RSY03).  There were seven participating stations: Effelsberg, Hartebeesthoek, Jodrell Bank (MK2), Medicina, Onsala, Torun, and Westerbork (single dish). We selected the source J1342$+$0504 as the phase-referencing calibrator. The nodding cycle time was about 4.5 min (1 min on the calibrator and 3.5 min on the targets). We also observed a bright source, 1156$+$295, as the bandpass calibrator. The experiment was correlated with the EVN software correlator \citep[SFXC,][]{kei15} at JIVE (Joint Institute for VLBI, ERIC) in real-time mode. Each station had a data transferring speed of 1024 Mbps (16 subbands, dual polarisation, 16~MHz per subband, 2~bit quantization). To image both target sources with a wide enough viewing field ($>1$ arcmin), we requested 1\,s integration time and 128 frequency points per subband.

\begin{table*}
\caption{Summary of the EVN imaging and circular Gaussian fitting results of CXO J133815.6$+$043255 and NGC~5252. Columns give (1) source name, (2 -- 5) J2000 positions and errors, (6) total flux density and $1\sigma$ error,  (7) source angular size, (8) brightness temperature and (9) peak brightness and $1\sigma$ formal error. } \label{tab1}
\setlength{\tabcolsep}{6pt}
\centering
\begin{tabular}{ccccccccc}
\hline
Source  Name  &  $\alpha$                                & $\sigma_\mathrm{\alpha}$
                                                                 & $\delta$                         & $\sigma_\mathrm{\delta}$
                                                                                                               &  $S_\mathrm{tot}$ & $\theta_\mathrm{size}$
                                                                                                                                              & $T_\mathrm{b}$ & $S_\mathrm{peak}$\\
              &   (J2000)                                & (mas) & (J2000)                          & (mas)    &   (mJy)          & (mas)     & ($10^{8}$~K)   & (mJy\,beam$^{-1}$)\\
\hline
CXO J133815.6$+$043255
              &   $13^\mathrm{h}38^\mathrm{m}15\fs63924$ & 1.1   &  $+04\degr32\arcmin55\farcs3636$ & 1.2      & 1.8$\pm$0.1      & $\le$4.0  &  $\ge$0.5     & 1.10$\pm0.05$ \\
NGC~5252      &   $13^\mathrm{h}38^\mathrm{m}15\fs87082$ & 0.9   &  $+04\degr32\arcmin33\farcs5184$ & 0.9      & 3.6$\pm$0.2      & $\le$3.9  &  $\ge$1.2     & 2.2$\pm0.1$ \\
\hline
\end{tabular}
\end{table*}

The visibility data were calibrated in the NRAO package Astronomical Imaging Processing Software \citep[AIPS, ][]{gre03}. The amplitude calibration was done with the system temperatures measured during the observations and the antenna gain curves. The apparent source position of CXO J133815.6$+$043255 was recalculated using the AIPS task CLCOR by adding a tiny position offset twice, with different signs. Note that the recalculation is a necessary step in the case of treating EVN data with AIPS to shift the image centre from CXO J133815.6$+$043255 to NGC 5252. The ionospheric delay was calibrated by the maps of total electron content. Parallactic angle correction was also applied. We removed the instrument phase error across subbands first, and then did a global fringe-fitting combining all the subbands. Only the phase bandpass solutions were derived via the bright calibrator 1156$+$295. We first imaged the phase-referencing calibrator J1342$+$0504 in Difmap \citep{she94} and then re-ran the global fringe-fitting in AIPS to remove the phase error coming from its source structure. The centroid of the brightest component (most likely the radio core) was used as the reference origin. We also did both amplitude and phase self-calibration on the phase-referencing source J1342$+$0504 and transferred its solutions to the CXO J133815.6$+$043255 data set. The calibrator J1342$+$0504 has quite rich structure on scales up to kpc in the North-South direction. The elongated structure is also in agreement with its image obtained from the VLA survey FIRST \citep[Faint Images of the Radio Sky at Twenty-Centimeters,][]{bec95}.

We deconvolved the calibrated visibility data with Difmap. Our correlation phase centre was at the position of CXO J133815.6$+$043255. To image NGC~5252 without suffering bandwidth smearing effect, we shifted the source position from CXO J133815.6$+$043255 to NGC 5252 in AIPS. We excluded the data from the shortest baseline of Effelsberg to Westerbork by setting the lower limit to 1.5 Mega-wavelength during the imaging process of the two sources because this baseline observed some very extended emission features, likely from the kpc relic jet of NGC~5252 \citep{the01}, which gave a baseline-dependent imaging error. We fitted the visibility data to a circular Gaussian model in Difmap and reported the fitting results in Table~\ref{tab1}. The position error was estimated via $\frac{\mathrm{FWHM}}{2\mathrm{SNR}}$, where FWHM is full width at half magnitude of the beam and SNR is signal to noise ratio. Note that we also included a systematic error of 5\,\% in the 1$\sigma$ error of the total flux density.

\section{EVN detection of CXO J133815.6$+$043255}
\label{sec3}

The final EVN imaging results are shown in the left two panels of Figure~\ref{fig1}. Both CXO J133815.6$+$043255 and NGC~5252 are clearly detected. Comparing with the {\it Chandra} X-ray image, the EVN images have a much higher resolution, up to 3~mas in the North-South direction.

In the image of natural weighting, we detected CXO J133815.6$+$043255 with a SNR of about 20. Our circular Gaussian model fitting gives a total flux density of $S_\mathrm{tot}=1.8\pm0.1$~mJy. The EVN position is within the 1$\sigma$ error circle of the VLA position \citep{kuk95}.

NGC~5252 has a total flux density of $3.6\pm0.2$~mJy, about a factor of 2 lower than that observed by \citet{the01} with the MERLIN at 1.6~GHz on 1995 December 3. This is expected, since NGC~5252 has some diffuse jet emission on the scale of a kpc. Actually, we also found a hint for these missing emission features. There are some regular patterns with peaks at $6\sigma$ level in the residual map, but, we failed to locate these extended features reliably because of the limited $uv$ coverage.

Thanks to the long baselines to Hartebeesthoek, the two EVN images have about 10 times better resolution ($\sim$3 mas) in North-South. We also tried fitting a point source model to the two data sets. The fitting results were not acceptable on the long baselines to Hartebeesthoek. The partially resolved structure, we think, is likely due to an instrumental effect. On one hand, there were some residual phase errors mainly coming from the un-modelled ionospheric delay. On the other hand, there are some weak and extended emission features that were missed in the EVN images. Considering the above issues, the size measurements are most likely dominated by the systematic error. These numbers listed in Table~\ref{tab1} should be taken as upper limits. A circular beam with a FWHM of 28 mas was also used in the position error estimation.

We also estimated the brightness temperature via the following equation \citep[e.g.,][]{an13},
\begin{equation}
T_\mathrm{b}=1.22\times10^{9}(1+z)\frac{S_\mathrm{tot}}{\nu_\mathrm{obs}^2\theta_\mathrm{size}^2}~\mathrm{K},
\end{equation}
where $S_\mathrm{tot}$ is the total flux density in mJy, $\theta_\mathrm{size}$ is the diameter of the circular Gaussian model in mas, $\nu_\mathrm{obs}$ is the observing frequency in GHz, and $z$ is the redshift. The results are listed in the last column of Table~\ref{tab1}. According to our EVN imaging results, the compact component in CXO J133815.6$+$043255 has a brightness temperature $T_\mathrm{b}\geq5\times10^{7}$~K. Thus, the origin of the radio emission is non-thermal synchrotron radiation.

\section{Discussion}
\label{sec4}

\subsection{Radio core in CXO J133815.6$+$043255}
\label{sec4-1}
All the previous flux density measurements of the ULX are summarised in Table~\ref{tab2}. The flux density measurement of the EVN is consistent with that of MERLIN. However, the two measurements are significantly lower than that observed by the VLA at the same observing frequency. On the other side, all the VLA measurements at $\sim$1.5 or 8.4 GHz agree with each other within their 3$\sigma$ error bars. The X-ray luminosity did not show significant variability either on a time scale of $<10$ yr \citep{kim15}. Thus, the ULX most likely has a relatively stable radio luminosity on a time scale of years.  The flux density loss of the MERLIN and EVN images is because their very high resolution completely resolves out some extended emission regions. The missing radio emission may be associated with relic jets \citep{dad10} or nearby star-forming activity.

Given that there is no significant luminosity variability in the X-ray and radio bands, we think that the long-lived compact component is a steady radio core instead of a rapidly expanding plasma ejecta (i.e. a transient jet). Transient jets are often seen during outbursts in X-ray BH binaries \citep[e.g.,][]{yan11} but rarely observed in ULXs \citep[e.g.,][]{cseh15b}.

The simultaneous multi-frequency VLA observations by \citet{wil94} gave a spectral index of $\alpha\approx -0.5$ ($S_\nu \propto \nu^\alpha$) between 1.4 and 8.4~GHz. Considering that the VLA observations might include some extended radio emission at 1.4 and 5 GHz, the spectral index measurement should be taken as a lower limit. The average flux density is 1.5~mJy at 8.4~GHz (Table~\ref{tab2}). Since the VLA images at 8.4 GHz had a sub-arcsecond resolution, similar to that of the MERLIN image at 1.6 GHz, the emission comes from the same region and therefore these data can be used to obtain a more realistic estimate for the spectral index, $\alpha\approx -0.3$.  Comparing with the average flux density and assuming no significant long-term variability, the VLBI-detected component has a spectral index of $\alpha\approx -0.1$ between 1.6 and 8.4 GHz, i.e. a flat radio spectrum.

In view that the EVN-detected component has a high brightness temperature, a partially optically thick radio spectrum, and a relatively stable luminosity, we identify it as the radio core, i.e. the jet-launching base, of CXO J133815.6$+$043255.

\subsection{Dual radio-emitting AGN}
\label{sec4-2}

By analogy with NGC 3341B \citep{bia13}, CXO J133815.6$+$043255 is also likely an LLAGN sitting in a dwarf galaxy \citep{kim15}. Its [O $\mathrm{III}$] and H$\mathrm{\alpha}$ luminosities are, respectively, $10^{39.7}$ and $10^{39.2}$ erg~s$^{-1}$ \citep{kim15}, comparable to that of nearby LLAGNs \citep{ho03}. The [O $\mathrm{III}$] luminosity is also correlated with both the X-ray and radio luminosities in a way similar to other LLAGNs \citep{kim15}. If CXO J133815.6$+$043255 is an LLAGN, the detection of a compact radio core is a high-probability event. The detection fraction of radio cores in the nearby LLAGNs is quite high, $\sim$30\% by the VLA at 15 GHz in Seyfert 2 galaxies \citep{ho08} and $\sim$100\% by the EVN at 1.4 GHz in some early-type galaxies of the Perseus cluster \citep{par16}.

The central object in CXO J133815.6$+$043255 is most likely a supermassive BH. The mass of a BH can be estimated via the BH fundamental plane relation \citep[e.g.,][]{mil12}:
\begin{eqnarray}
\nonumber
\log M_\mathrm{BH}=(1.638\pm0.070)\log L_\mathrm{R} \\
  -(1.136\pm0.077)\log L_\mathrm{X}-(6.863\pm0.790),
\end{eqnarray}
where $L_\mathrm{X}$ is the 2--10 keV X-ray luminosity, $L_\mathrm{R}=\nu L_\mathrm{\nu}$ is the 5 GHz radio luminosity (both in erg\,s$^{-1}$), and $M_\mathrm{BH}$ is the BH mass in M$_\odot$, with an empirical uncertainty of 0.44 dex ($1\sigma$). This leads to a mass of $10^{9.1\pm1.2}\mathrm{M}_\odot$. Even allowing for a large uncertainty of up to 5$\sigma$, it is still hard to classify the source as an intermediate-mass BH. The fundamental plane relation is only applicable to BHs accreting at significantly sub-Eddington rates. In the low-accretion rate state, the radio emission is dominated by partially synchrotron self-absorbed compact jets that have flat spectra and show relatively steady structure on mas scales. In case of the source, the EVN detection of the radio core provides the direct observational confirmation of the disc-jet coupling. On the other hand, most ULX radio counterparts have a nebula morphology on the scale of arcsecond \citep[e.g.,][]{cseh12}. As far as we know, a compact and steady component on VLBI scales in the hard state has not been presented previously in any off-nucleus ULX. With the EVN observations, \citet{mez15} reported a questionable weak jet component in ULX NGC~2276-3c at 1.6 GHz. We did an independent reanalysis of the EVN data and failed to confirm the detection. \citet{mez14} observed another ULX, NGC 4088 X1, while, they were unsuccessful in achieving a reliable (SNR$\geq$6) detection and identifying it as a synchrotron emission jet ($T_\mathrm{b}\geq10^6$\,K).

Considering all the above properties, CXO J133815.6$+$043255 can be naturally explained as an LLAGN. It is the first example of a supermassive BH identified with an ULX. \citet{Mapelli12} considered a minor merger scenario for HLX--1, but with a BH mass in the range $10^3-10^5\mathrm{M}_\odot$. \citet{Jonker10} proposed that CXO J122518.6$+$144545 is a recoil BH, but they did not provide an estimate of its mass. It is also reported that some ULX candidates turned out to be background AGNs \citep[e.g.,][]{Foschini02, Sutton15}. In case of CXO J133815.6$+$043255, it is unlikely to be a background AGN and the host galaxy may be a merging dwarf galaxy \citep{kim15}.

Since NGC 5252 also hosts a supermassive BH \citep{cap05}, CXO J133815.6$+$043255 and NGC 5252 are a promising dual radio-emitting AGN system with a projected separation of about 10~kpc. Currently, among the great number of optical candidates, there are only a few examples of binary AGN candidates confirmed in the radio with direct VLBI imaging observations, such as SDSS J1502$+$1115 \citep{dea14}, SDSS J1536$+$0441 \citep{bon10} and B0402$+$379 \citep{rod06}. Most of the optical binary AGN candidates only have only a single radio-emitting AGN \citep[e.g., the recent VLBI studies by][]{an13, bon16, fre12, gab16}. Our finding shows that VLBI observations of ULXs with compact radio counterparts are an effective tool to identify dual AGNs.

\begin{table}
\caption{List of the previous radio observations of the X-ray source CXO J133815.6$+$043255.} \label{tab2}
\footnotesize
\setlength{\tabcolsep}{4pt}
\begin{tabular}{cccccc}
\hline
Date       &  Array         & $\nu_\mathrm{obs}$ & $S_\mathrm{tot}$  & Beam Size        & Ref.   \\
yyyy-mm-dd &   (mJy)        & (GHz)              & (mJy)             & ($\arcsec$)      &       \\
\hline

1990-04-29 &  VLA           & 1.5                & $3.2\pm0.2$       & $1.6\times1.3$   &  a    \\
1993-01-18 &  VLA           & 1.4                & $3.7\pm0.1$       & $1.6\times1.4$   &  b    \\
1995-12-03 &  MERLIN        & 1.6                & $2.2\pm0.1$       & $0.36\times0.29$ &  c    \\
2000-12-23 &  VLA           & 1.4                & $3.8\pm0.2$       & $7.1\times5.9$   &  d    \\
\hline
1990-04-29 &  VLA           & 8.4                & $1.4\pm0.2$       & $0.26\times0.21$ & a     \\
1991-06-24 &  VLA           & 8.4                & $1.6\pm0.1$       & $0.33\times0.24$ & e     \\
1992-04-28 &  VLA           & 8.4                & $1.5\pm0.1$       & $3.0\times2.4$   & e     \\
1993-01-18 &  VLA           & 8.4                & $1.4\pm0.1$       & $0.21\times0.23$ & b     \\
\hline
1990-04-29 &  VLA           & 4.9               & $1.9\pm0.2$       & $0.46\times0.38$ & a     \\
\hline
\end{tabular} \\
References: a -- \citet{wil94}, b -- \citet{nag99}, c -- \citet{the01}, d -- FIRST \citep{bec95}, e -- \citet{kuk95}.
\end{table}

\section{Conclusion}
\label{sec5}

The source CXO J133815.6$+$043255 shows clear counterparts in the UV, optical, and radio bands. EVN observations at 1.7 GHz reveal a single radio component with a total flux density $1.8\pm0.1$~mJy. In view of its parsec-scale compact radio structure and flat radio spectrum, we think that it is most likely a compact radio core associated with an AGN. Considering the optical and X-ray characteristics, and the mass constrained by the fundamental plane relation, we identify CXO J133815.6$+$043255 as a LLAGN with an accreting supermassive BH. Moreover, CXO J133815.6$+$043255 and NGC 5252 are a promising candidate of dual radio-emitting AGN system.

\section*{Acknowledgements}
\footnotesize
XL and XL are supported by the 973 Programme 2015CB857100; the Key Laboratory of Radio Astronomy, CAS; and the National Natural Science Foundation of China (No. 11273050). TA was supported by the SKA pre-construction funding from the China Ministry of Science and Technology under grant No. 2013CB837900, TA thanks the grant support by the Youth Innovation Promotion Association CAS. LCH acknowledges financial support from the Chinese Academy of Science through grant No. XDB09030102 (Emergence of Cosmological Structures) from the Strategic Priority Research Programme, and from the National Natural Science Foundation of China through grant No. 11473002. This work was supported by grant 2016YFA0400702 from the Ministry of Science and Technology of China. LC thanks the Programme of the Light in China's Western Region (Grant No. YBXM-2014-02) and the National Natural Science Foundation of China (Grant No. 11503072). SB and ZP are members of the MAGNA project\footnote{http://www.issibern.ch/teams/agnactivity/Home.html}. The European VLBI Network is a joint facility of independent European, African, Asian, and North American radio astronomy institutes. Scientific results from data presented in this publication are derived from the following EVN project code: RSY03. The research leading to these results has received funding from the European Commission Seventh Framework Programme (FP/2007-2013) under grant agreement No. 283393 (RadioNet3). e-VLBI research infrastructure in Europe is supported by the European Union's Seventh Framework Programme (FP7/2007-2013) under grant agreement number RI-261525 NEXPReS. This research has made use of data obtained from the Chandra Data Archive and the Chandra Source Catalog.









\bsp	
\label{lastpage}
\end{document}